\documentclass[Crown,sagev,times]{sagej}

\usepackage{moreverb,url}
\usepackage{mathtools}
\usepackage{subcaption}

\usepackage[colorlinks,bookmarksopen,bookmarksnumbered,citecolor=red,urlcolor=red]{hyperref}

\newcommand\BibTeX{{\rmfamily B\kern-.05em \textsc{i\kern-.025em b}\kern-.08em
T\kern-.1667em\lower.7ex\hbox{E}\kern-.125emX}}

\def\volumeyear{2022}

\begin{document}

\runninghead{Bonavia}

\title{Viscoelastic Material Properties of Gelatin with Varying Water to Collagen Mass Ratios}

\author{J. E. Bonavia\affilnum{1}}

\affiliation{\affilnum{1}Massachusetts Institute of Technology, Department of Mechanical Engineering, Cambridge MA, USA}

\corrauth{Joseph E. Bonavia, \\
Massachusetts Institute of Technology,
Department of Mechanical Engineering, \\
77 Massachusetts Avenue, 
Cambridge MA,
02139, USA.}

\email{jbonavia@mit.edu}

\begin{abstract}
Gelatin is often used as an analog for studying soft and biological materials in order to understand the mechanics of behavior of biological tissue in events like traumatic brain injuries. The material properties of gelatin change with the ratio of water to gelatin powder used to make a given sample. Characterizing the relationship between this ratio and the material properties of gelatin is crucial to enable its use in mechanics experiments. In this work, compression tests were performed on a texture analyzer on samples which ranged from a 2:1 to 20:1 ratio of water to gelatin powder. In this range, instantaneous stiffnesses were well fit via power law in this ratio and decreased from 277 ± 30 kPa to 4.34 ± 0.64 kPa. The dominant (longest) timescales of the samples were well fit via a sigmoid function in this ratio and increased from 29.8 ± 1.0 s to 621 ± 92 s. The resulting ratio-property relationships offer a functional way to design gelatin samples for use in mechanics experiments. 
\end{abstract}


\maketitle
\section{Introduction and Background}
\subsection{Introduction}
In recent years, research into soft-matter mechanics has become increasingly important as scientists and engineers attempt to understand the underlying physics of injuries and diseases and improve the design of medical devices and treatments. When performing any experiment, it is essential that your test system is well characterized. However, because soft material mechanics is a relatively new field, soft materials are generally highly complex and non-linear in nature and soft materials are notoriously difficult to characterize, there are few standardized model materials for use in soft-matter mechanics experiments.

Gelatin, commonly used as an ingredient in the kitchen, is made up of collagen; the same long, yarn-like molecule which makes up our skin and bones. It is derived from the bones of pigs, sheep, and cows which makes it cheap and accessible \cite{Cite1}. Ballistics researchers often manufacture gelatin with properties representative of human flesh when testing projectiles and body armor (generally with a ratio of 8-10 parts water to 1 part powder by mass) \cite{Cite1,Cite2}. It is also translucent which makes it easy to see internal damage. Finally, manufacturing a gelatin sample is safe and easy since it requires only an edible collagen powder and water, and it releases easily from molds. By varying the amount of powder and water used, samples can be made stiffer and stronger, or softer and weaker. This makes gelatin an ideal analog for studying soft, “squishy,” biological materials like those that make up most of the human body. 

Gelatin is safer, cheaper, and cleaner to manufacture in a lab than many other common soft materials (like PDMS and synthetic hydrogels which are messy and whose manufacturing processes involve toxic chemicals). Because it is a relatively ubiquitous material, it also offers a unique opportunity to act as a kind of “standard” in experiments where it is representative. However, for a material to be useful when using it as an analog to understand some phenomenon, it is important that it is well characterized. This means that it is important that the relationship between strain, strain-rate, and stress is well understood. The elastic moduli ($E_0$, $E_1$, $E_2$, $... \,E_n$) of a material describe the relationship between strain and stress, and the relaxation times of a material ($T_1$, $T_2$, $...\,T_n$) describe the time scales over which these stresses relax \cite{Cite3}. To characterize gelatin at varying mass-ratios of water to collagen powder, stress-relaxation experiments were performed on a Texture Analyzer, and a linear viscoelastic model was then fitted to the resulting data. 

The samples were prepared according to the instructions on the package, with varying ratios of water to gelatin powder (2:1, 4:1, 8:1, 12:1, 16:1, and 20:1) by mass. The upper and lower bounds of the experiments were determined by which ratios could be successfully manufactured and tested with simple tools. The samples were then compressed and held at constant strain in the Texture Analyzer, which measured the decrease in stress over time resulting from the viscoelasticity in the material. A viscoelastic model consisting of three elastic moduli and two viscosities (the form of which is related to the hierarchical physical structure of gelatin) was then fitted to the data and the moduli and time constants were then extracted from the model. The resulting parameters were compared with existing literature on collagen and gelatin material parameters for general accuracy. Finally, the moduli and time constants were plotted against the water to collagen ratio and curve to determine an underlying relationship between water content and viscoelastic properties. The resulting water content - property relationships enable the use of gelatin as a representative material in soft-matter physics research. 

\subsection{Background}
\begin{figure}[htbp]
\centering
\begin{subfigure}[t]{0.48\linewidth}
    \centering
    \includegraphics[width=\linewidth]{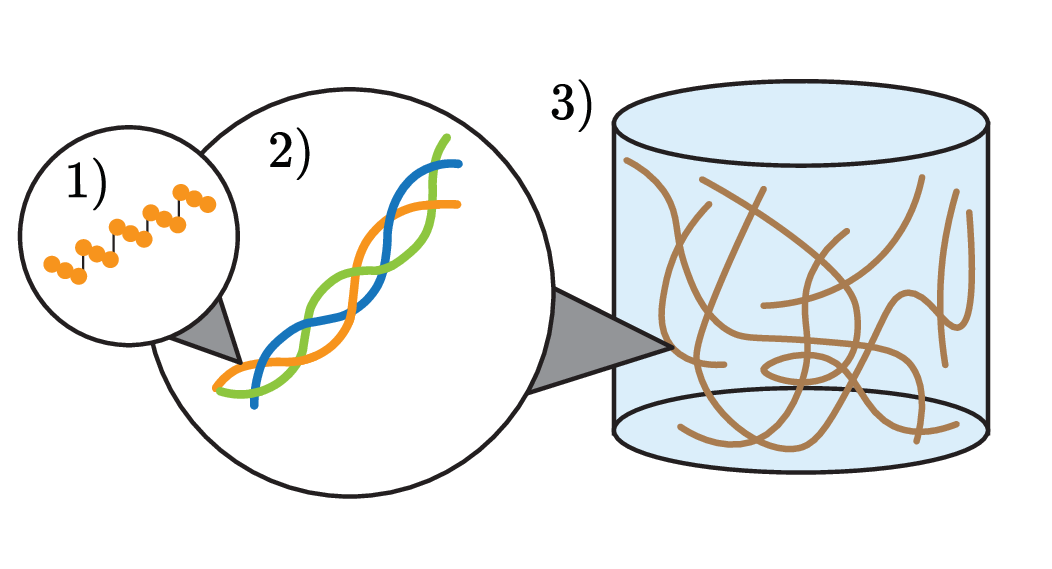}
    \caption{}
    \label{fig:gelatin_micro}
\end{subfigure}
\hfill
\begin{subfigure}[t]{0.48\linewidth}
    \centering
    \includegraphics[width=\linewidth]{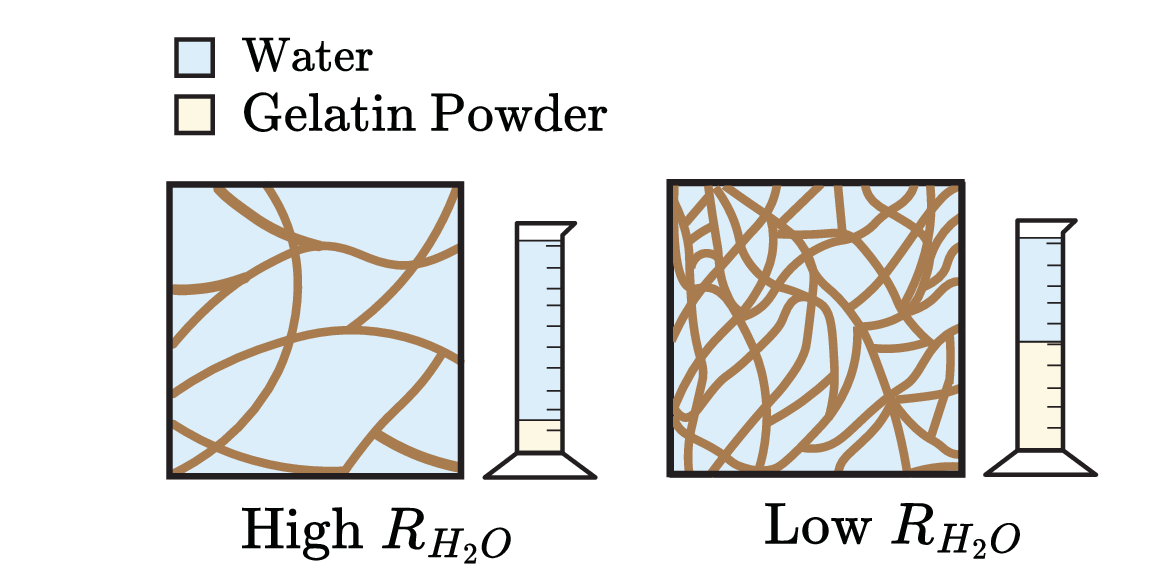}
    \caption{}
    \label{fig:gelatin_macro}
\end{subfigure}

\caption{(a) The hierarchical structure of gelatin. 1) Long amino acid chains are the base units of collagen. 2) The long chains are ``woven'' together into yarn-like collagen strands. 3) The strands are tangled and linked together, capturing water and forming gelatin. 
(b) Relative density of polymer networks for high and low water to collagen mass ratio $R_{H_2O}$. The graduated cylinders depict the relative amounts of water and gelatin powder (collagen) used.}
\label{fig:gelatin_structure}
\end{figure}

Gelatin is a hydrogel, meaning that it consists of an interconnected network of long organic molecules called polymers that trap and hold large mass fractions of water. In gelatin, the polymer is collagen, a complex protein, whose long chains are “woven” together to form a yarn-like structure (Fig. \ref{fig:gelatin_micro}) \cite{Cite4}. When the collagen powder is dissolved in hot water, short strands of collagen are separated from each other, and when the mixture cools, the strands are tangled and linked together, trapping water between the chains. By changing the ratio between the water and gelatin powder used to make the sample, the density of the polymer network within the polymer is also changed. Water to collagen mass ratio $R_{H_2 O}$ is  defined as the mass of the water used to make the sample divided by the mass of gelatin powder used to make the sample. As $R_{H_2 O}$ increases (increasing water content), the polymer network gets less dense (Fig. \ref{fig:gelatin_macro}).

\begin{figure}[htbp]
\centering
\includegraphics[width=0.6\linewidth]{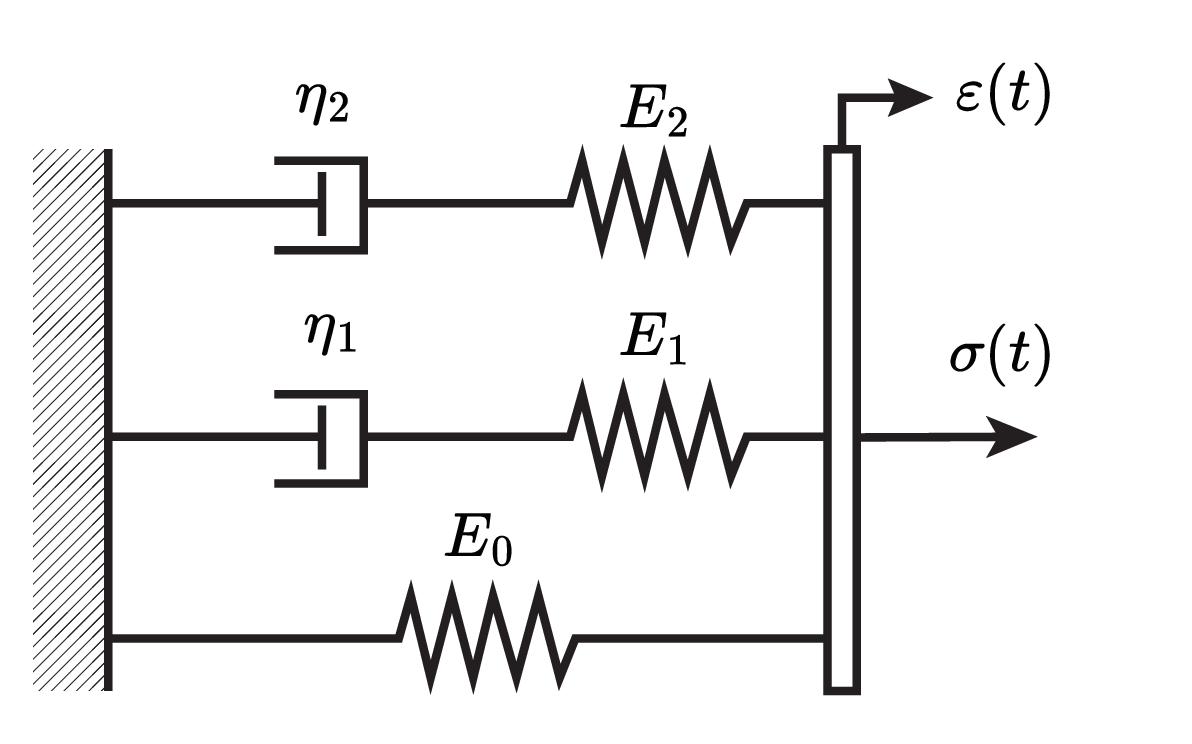}
\caption{Maxwell--Weichert spring--damper model of gelatin. Initially during a step response, the viscous dampers react rigidly and the system responds with the instantaneous modulus $(E_0 + E_1 + E_2)$. As $t\to\infty$, the viscous dampers support none of the load, and the relaxation modulus decays to the quasi-static modulus $E_0$.}
\label{fig:maxwell_weichert_model}
\end{figure}

As the gelatin is stretched and squished, these collagen molecules slide against each other leading to energy loss and in turn to a phenomenon known as viscoelasticity \cite{Cite3}. This means that the stress $\sigma(t)$ [Pa] is not only dependent on the strain $\varepsilon(t)$ [-] but also on the strain rate $\dot\varepsilon(t)$ [1/s]. Water flows between the chains generating additional viscous losses in a phenomenon known as nano-poroelasticity, which manifests itself within the viscoelastic response \cite{Cite4}. The relationship between the stress and strain in a collagen-based gel can be modeled through the use of the Maxwell–Weichert spring-damper system (Fig. \ref{fig:maxwell_weichert_model}) with three elastic moduli ($E_0$, $E_1$, and $E_2$) [Pa] and two viscosities ($\eta_1$ and $\eta_2$) [Pa$\cdot$s]\cite{Cite3}. The relaxation modulus response $E_r(t)$ [Pa] of this model to a step input axial strain of magnitude $\varepsilon_0$ is given as:
\begin{equation}\label{eq:MaxWei}
E_r(t) \equiv \frac{\sigma(t)}{\varepsilon_0}= E_0+E_1\exp\left(\frac{-t}{T_1}\right)+E_2\exp\left(\frac{-t}{T_2}\right)
\end{equation}
where the time constants (also referred to as relaxation times) $T_1$ and $T_2$ represent the e-folding times of the two parallel stiffnesses of the spring damper model $T_i \equiv \eta_i/E_i$ [s].

For gelatin and other collagen-based hydrogels, there tends to be one long and one short time scale, which can differ by several order of magnitudes \cite{Cite5,Cite6}. During a perfect step strain input, the strain rate is initially infinite. This means that the viscous dampers in the Fig. \ref{fig:maxwell_weichert_model} act as rigid bodies and the relaxation modulus at time equals zero is the sum of the three elastic moduli. This initial relaxation modulus is known as the instantaneous modulus ($E_0+E_1+E_2$) of the material. As time increases the effective stiffness of the sample decreases exponentially at a rate characterized by the time constants ($T_1$ and $T_2$) of the system. This is due to the untangling of the chains and redistribution of water within the sample. As time goes to infinity the relaxation reaches a set value known as the quasi-static modulus $E_0$.

Several previous studies have been performed on ballistic gelatin and other collagen-based hydrogels using indentation tests. These tests use a small probe to press into a gelatin sample and measure the force vs. distance the probe has pressed into the surface of the material. The data from these tests can then be fitted to a model of the deformation of viscoelastic half space to retrieve the relevant parameters elastic moduli and viscosities. A 2013 study on collagen fibrils found elastic moduli several orders of magnitude higher than those expected of gelatin gels \cite{Cite6}. This is reasonable, as the collagen strands consist of much longer, thoroughly cross-linked chains than those in gelatin. However, the study also found relaxation times of 7 ± 2 sec and 102 ± 5 sec for the short and long timescales respectively, which are quite similar to values found here for low water ratio gelatin. A 2021 study using machine learning methods to fit indentation test data focused on only the longer of the two viscoelastic timescales and found it to be approximately 131 seconds \cite{Cite7}. Finally, another more thorough study from 2021 measured the viscoelastic properties of gelatin-based hydrogels at varying water contents and found that a mass-water ratio of 9:1, 2.3:1, and 1:1 gelatin samples had a stiffnesses of 2.64 kPa, 9.75 kPa, and 27.74 kPa respectively. It also found that with increased water content, gelatin becomes more elastic \cite{Cite8}. However, experiments in this prior work were performed at 27 °C (roughly 7 °C higher temperature) and were prepared using a solvent casting method which may explain significant differences to the results of this experiment. 
\section{Experimental Design}
To attain accuracy and precision of results, procedures were put in place during both sample preparation and running of the experiments in order to keep tight control over any extraneous variables. During sample preparation, following this procedure helped to tightly control the water ratio and geometry of the samples. During the experiments, following the procedure helped to ensure that measurements of length, force, and temperature were precise and accurate so that analysis of the resulting data would be useful in interpreting any relevant trends. 
\subsection{Sample Preparation}
The preparation of gelatin samples was based on the instructions provided by the manufacturer on a jar of Knox gelatin\ powder (with the exception of the ratios of gelatin powder and water used). All measurements of mass were performed using the AccuWeight IC255 Digital Scale which has an error of ± 0.05\%. Water was boiled in the Comfee CEKS001 Water Kettle. The refrigerator temperature was relatively constant at roughly 2 ± 2 $^\circ$C. Tap water (from Cambridge, Massachusetts) was used in both the room temperature and hot water steps, and any concentrations of impurities in it are assumed to be negligible. Each sample was prepared using a total of 120 g of water. This is because the 3D printed molds had a volume slightly lower than 120 mL. The necessary mass of gelatin powder was then calculated based on the desired mass ratio of water to gelatin powder (2:1, 4:1, 8:1, 12:1, 16:1, and 20:1). The steps used to prepare the gelatin samples are as follows:
\begin{enumerate}
    \item Place 250 mL plastic beaker on scale. Tare scale.
    \item Add the desired mass of gelatin powder to the beaker.
    \item Tare again. Add 60 g (half the total amount of water) of room temperature water to the beaker. 
    \item Stir together with a plastic spoon for two minutes. The resulting mixture will be a gelatinous paste. 
    \item Tare again (leaving the spoon in the beaker). Add 60 g of boiling water to the beaker mix until fully dissolved. It may be necessary to place the beaker in a bowl of hot water to keep everything at temperature.
    \item Pour gelatin into molds and place into refrigerator for 4 hours. 
    \item Let samples sit for about 8 hours or until slightly colder than 20 $^\circ$C.
    
\end{enumerate}
This procedure was repeated for each of the desired mass ratios and resulting “actual” mass ratios are given in Table \ref{tab:mass_ratios}. 
\begin{table}[htbp]
\centering
\caption{The actual measured water-to-powder mass ratios of each gelatin sample.}
\label{tab:mass_ratios}
\begin{tabular}{|c|c|}
\hline
\textbf{Sample} & \textbf{Actual Water to Collagen Mass Ratio, $\boldsymbol{R_{H_2O}}$ [-]} \\
\hline
\hline
2:1  & $2.012 \pm 0.051$ \\
\hline
4:1  & $4.053 \pm 0.043$ \\
\hline
8:1  & $7.959 \pm 0.085$ \\
\hline
12:1 & $11.95 \pm 0.08$  \\
\hline
16:1 & $16.02 \pm 0.10$  \\
\hline
20:1 & $19.98 \pm 0.10$  \\
\hline
\end{tabular}
\end{table}
\subsection{Experimental Setup and Procedure}
Before testing each sample, they were first measured for their dimensions and temperature. Because the samples varied slightly in height and diameter across their cross sections, five measurements were taken in different locations, and the mean value and error were calculated for the height and diameter of each sample. These measurements were performed using the Mitutoyo 2006 Ruler (± 0.5 mm). The temperature of each sample was then measured using the Etekcity 630 Infrared Thermometer (± 2 $^\circ$C). Again, five measurements were performed and averaged, this time because of the large error of the infrared thermometer. 

\begin{figure}[htbp]
\centering
\includegraphics[width=0.85\linewidth]{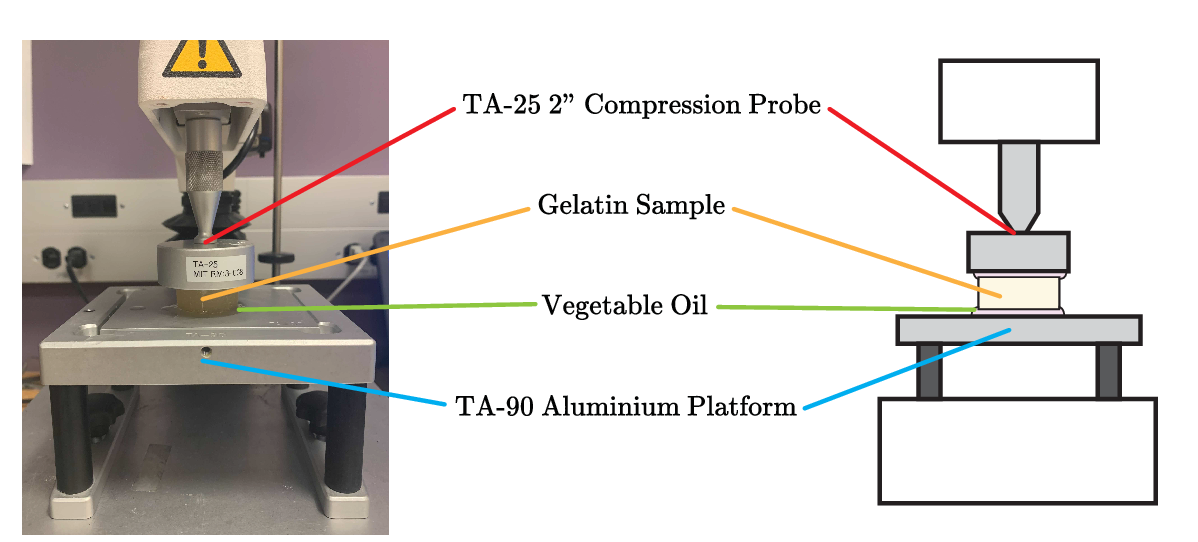}
\caption{Texture Analyzer compression experiment setup. Samples were lubricated with vegetable oil to mitigate sample barreling (which would increase error). They were then placed on the TA-90 Aluminum Test Platform and compressed via the TA-25 2-inch Compression Probe at 1 mm/s.}
\label{fig:compression_setup}
\end{figure}

Compression experiments were performed on the Texture Analyzer (TA). Fig. \ref{fig:compression_setup}, shows the setup for each run of the compression experiment. The sample was placed onto the TA-90 Aluminium test platform, and during each experiment, the TA-25 2 inch (5.08 cm) compression probe is lowered (at a rate of 1 mm/s) compressing the sample to approximately 15\% strain. At this strain rate the sample experiences what is essentially a step input in strain, with approximately 2.6\% error in stress due to the finite strain rate applied by the Texture Analyzer (explained further in discussion section). The sample was lubricated on its top and bottom surfaces with vegetable oil in order to mitigate the effect of barreling. Before each experiment, the TA was calibrated with a 2 kg weight. The probe was calibrated to be 30 ± 0.5 mm above the test platform (using the ruler), and once the samples were measured, a distance trigger was used to initiate the start of the test. This was found to be the most consistent way to control the compression amplitude during initial tests.

\begin{figure}[htbp]
\centering
\includegraphics[width=0.8\linewidth]{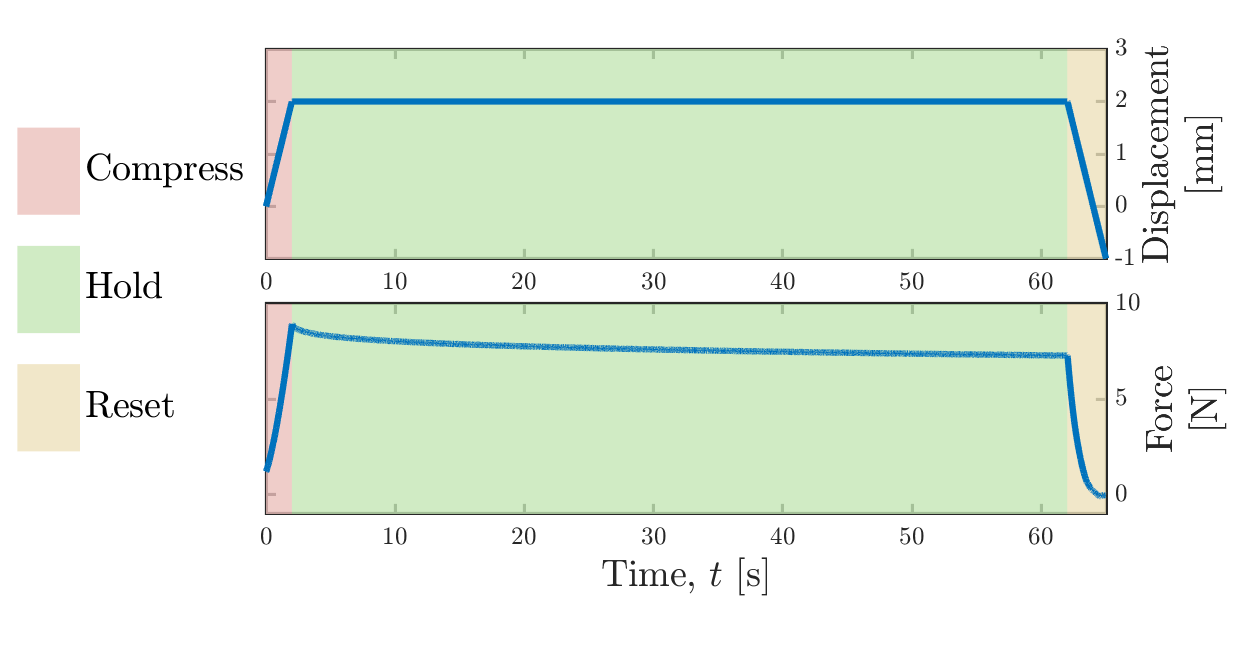}
\caption{A plot of a typical force and displacement curve for a gelatin sample in compression on the texture analyzer. During the ``compress'' stage (denoted in red) the crosshead moves down at a constant rate compressing the sample. During the ``hold'' stage (denoted in green) the sample is held at a constant strain. The stress response decreases due to viscoelastic effects. Finally, during the ``reset'' stage (denoted in yellow) the probe moves back up from the sample to its starting position.}
\label{fig:steps}
\end{figure}

The data from each experiment (as shown in Fig. \ref{fig:steps}) on the TA consisted of three main stages. First, during the “compression” stage, the probe was lowered until the sample was compressed to a prescribed strain. Force increases towards a maximum during this stage. Next, during the “hold” stage, the sample is held at constant strain. Due to viscoelasticity the force decreases exponentially during this stage. Further analysis is performed on this step. Finally, during the “reset” stage the force decreases as the probe is pulled away from the sample. 

\section{Results and Discussion}
Force-time data from the Texture Analyzer was converted into relaxation modulus-time data and then fit to (\ref{eq:MaxWei}). The elastic moduli and time constants were then extracted and plotted against $R_{H_2 O}$. The relationships between elastic moduli and water to collagen ratio were well fit via a decreasing power law, and the relationships between time constants  and water to collagen ratio were well fit via a sigmoid function. The stiffness range (which spans approximately two orders of magnitude) and resulting water content-gelatin property relationships enable the use of gelatin as a model material in soft matter mechanics research.
\subsection{“Step” Response Fitting}
The data from the hold stage of the “step” stress-relaxation experiment from the TA can be converted to measured relaxation modulus by dividing first by the measured cross-sectional area of the sample (resulting in the compressive stress) and then dividing by the constant value of strain during the “hold” stage. Next, (\ref{eq:MaxWei}) is fit to the resulting relaxation modulus vs. time curve, and the fit parameters (which are also the viscoelastic properties $E_0$, $E_1$, $E_2$, $T_1$, and $T_2$) of an individual sample can be extracted. Fig. \ref{fig:relaxation_unnormalized} shows plots of the relaxation modulus vs. time for one sample of each included mass ratio. This represents the relative magnitude of each response. Fig. \ref{fig:relaxation_normalized} shows the relaxation modulus normalized by the quasi-static modulus $E_0$, the value to which it decays. This allows us to observe the relative magnitude of the instantaneous modulus (the peak of the curve), and the differences in relaxation times (based on how fast the responses decay). Dashed lines represent extrapolated data based on the fit curve. Table \ref{tab:maxwell_weichert_params} contains the average values of the parameters for each set of samples. 
\begin{table}[htbp]
\centering
\caption{Average values of the Maxwell--Weichert viscoelastic material parameters for each set of samples.}
\label{tab:maxwell_weichert_params}
\begin{tabular}{|c||c|c|c|c|c|}
\hline
\textbf{Sample} & $\boldsymbol{E_0}$ \textbf{[kPa]} & $\boldsymbol{E_1}$ \textbf{[kPa]} & $\boldsymbol{E_2}$ \textbf{[kPa]} & $\boldsymbol{T_1}$ \textbf{[s]} & $\boldsymbol{T_2}$ \textbf{[s]} \\
\hline
\hline
\textbf{2:1}  & $230 \pm 23$   & $35.5 \pm 4.9$ & $11.3 \pm 2.3$  & $29.8 \pm 1.0$ & $2.04 \pm 0.11$ \\
\hline
\textbf{4:1}  & $59.9 \pm 7.0$ & $9.5 \pm 1.3$  & $3.49 \pm 0.19$ & $34.2 \pm 3.7$ & $2.42 \pm 0.29$ \\
\hline
\textbf{8:1}  & $27.7 \pm 6.9$ & $4.4 \pm 1.2$  & $1.00 \pm 0.31$ & $58 \pm 16$    & $3.73 \pm 0.31$ \\
\hline
\textbf{12:1} & $14.2 \pm 2.1$ & $4.60 \pm 1.10$& $0.678 \pm 0.099$ & $287 \pm 24$ & $15.8 \pm 1.8$ \\
\hline
\textbf{16:1} & $7.70 \pm 1.30$& $2.94 \pm 0.46$& $0.48 \pm 0.12$ & $576 \pm 85$   & $23.9 \pm 4.1$ \\
\hline
\textbf{20:1} & $2.83 \pm 0.43$& $1.32 \pm 0.14$& $0.179 \pm 0.072$ & $621 \pm 93$ & $21 \pm 10$ \\
\hline
\end{tabular}
\end{table}

\subsection{Viscoelastic Stiffnesses vs. Water-to-Collagen Mass Ratio}
Visually, the step response curves reveal some pretty self-evident trends in the way in which changes in the mechanical response of gelatin changes with the water to collagen mass ratio. In order to more quantitatively understand how this ratio affects the mechanical properties of the gelatin, values of the elastic moduli each sample are plotted versus the mass ratio of said sample. As shown in Fig. \ref{fig:moduli}, the moduli relationships are well fit via a decreasing power law of the form:
\begin{equation}\label{eq:moduli}
    E_i(R_{H_2O}) = a_i(R_{H_2O})^{-b_it}
\end{equation}
A weighted fit according to the section “Direct Estimation of Weights” in the NIST Engineering Statistics Handbook \cite{Cite9} was used to account for large differences in absolute error among the different test conditions. Table \ref{tab:moduli_fit_params} contains the values of the resulting parameters for all three moduli.

\begin{figure}[htbp]
\centering
\begin{subfigure}[t]{0.48\linewidth}
    \centering
    \includegraphics[width=\linewidth]{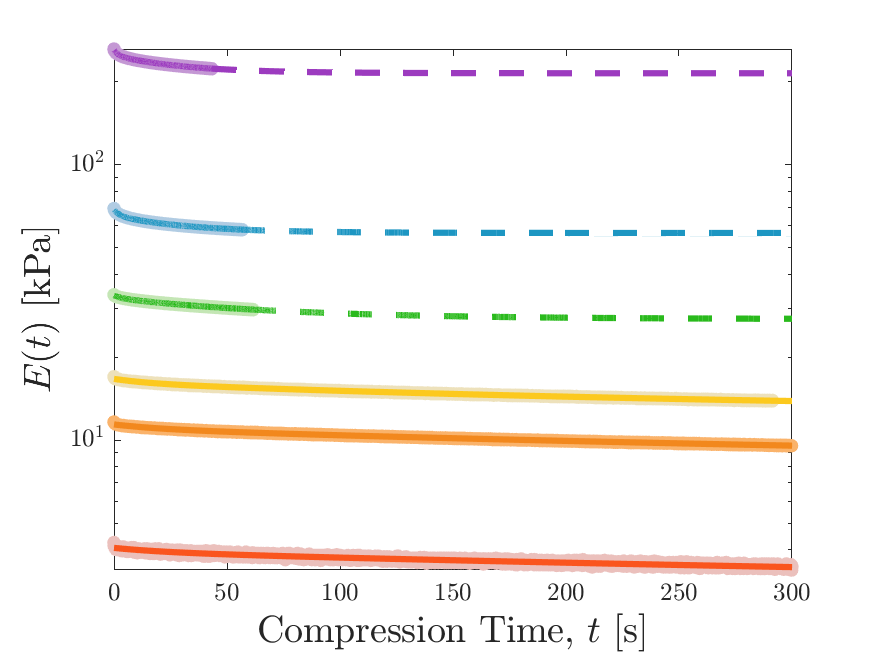}
    \caption{}
    \label{fig:relaxation_unnormalized}
\end{subfigure}
\hfill
\begin{subfigure}[t]{0.48\linewidth}
    \centering
    \includegraphics[width=\linewidth]{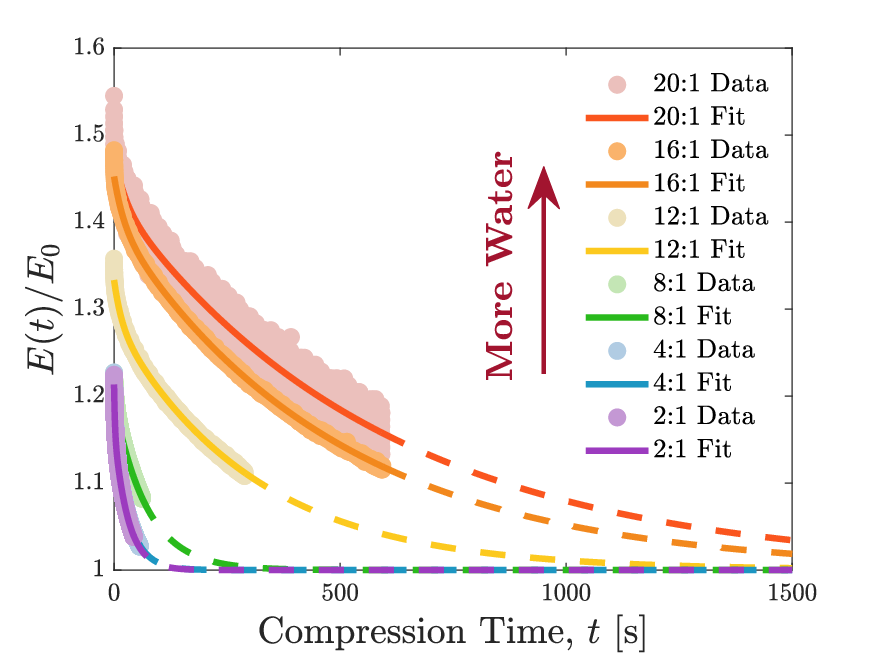}
    \caption{}
    \label{fig:relaxation_normalized}
\end{subfigure}

\caption{Relaxation modulus responses of one of each set of samples at roughly constant temperature ($19.9 \pm 0.5~^{\circ}\mathrm{C}$). 
(a) A semi-log plot of the relaxation modulus response of each sample to the step-response experiments. Values range over several orders of magnitude. 
(b) A plot of the normalized relaxation modulus with time. Notice the normalized curves seem to be bounded by a lower and upper limit.}
\label{fig:relaxation_modulus}
\end{figure}
In a hydrogel the elastic stiffness of the system comes largely from the stretching of the polymer (in this case collagen) chains. Therefore, it follows that increasing water ratio (and by extension decreasing collagen’s mass fraction) would decrease stiffness in the system. The resulting fit parameters (given in table \ref{tab:moduli_fit_params}) each have a positive value of b and therefore demonstrate this decreasing stiffness behavior.

\begin{table}[htbp]
\centering
\caption{Fit parameters of moduli vs.\ concentration. $a$ describes the hypothetical modulus of pure gelatin, and $b$ describes the power to which modulus decreases with the water ratio.}
\label{tab:moduli_fit_params}
\begin{tabular}{|c||c|c|}
\hline
\textbf{Modulus} & $\boldsymbol{a}$ \textbf{[kPa]} & $\boldsymbol{b}$ \textbf{[-]} \\
\hline
\hline
$\boldsymbol{E_0}$ & $656 \pm 87$  & $1.63 \pm 0.08$ \\
\hline
$\boldsymbol{E_1}$ & $52.8 \pm 17$ & $1.09 \pm 0.16$ \\
\hline
$\boldsymbol{E_2}$ & $30.4 \pm 6.6$ & $1.55 \pm 0.10$ \\
\hline
\end{tabular}
\end{table}

However, it is important to note that the ratio between the different elastic moduli is not as well fit by this model. As shown in Fig. \ref{fig:relaxation_normalized}, instantaneously at the moment the step response begins, the ratio between response and the final value of the relaxation modulus is not the same for all conditions. This value approaches a constant value at high and low mass-ratio. That is, the ratio between the instantaneous modulus ($E_0+E_1+E_2$) and quasi-static modulus $E_0$ is not constant for different concentration. This would imply that all three lines should be parallel in Fig. \ref{fig:moduli} at high and low concentrations but should deviate in between. This is indeed what we see from the data, however this proves a difficult set of data to fit with an elementary function (perhaps the product of a sigmoid and an exponential could work, but fitting such a function would be quite difficult). This effect is due largely to deviations in $E_1$ visible in Fig. \ref{fig:moduli}. It is quite possible that the effect of nanoporoelasticity within the matrix becomes more significant at higher water ratios, increasing the magnitude of the time dependent response. Within the experimental range, this deviation is relatively small and should not have a significant effect when using the power law to predict stiffness, and manufacture a sample at a specific water to collagen ratio. 
\begin{figure}[htbp]
\centering
\includegraphics[width=\linewidth]{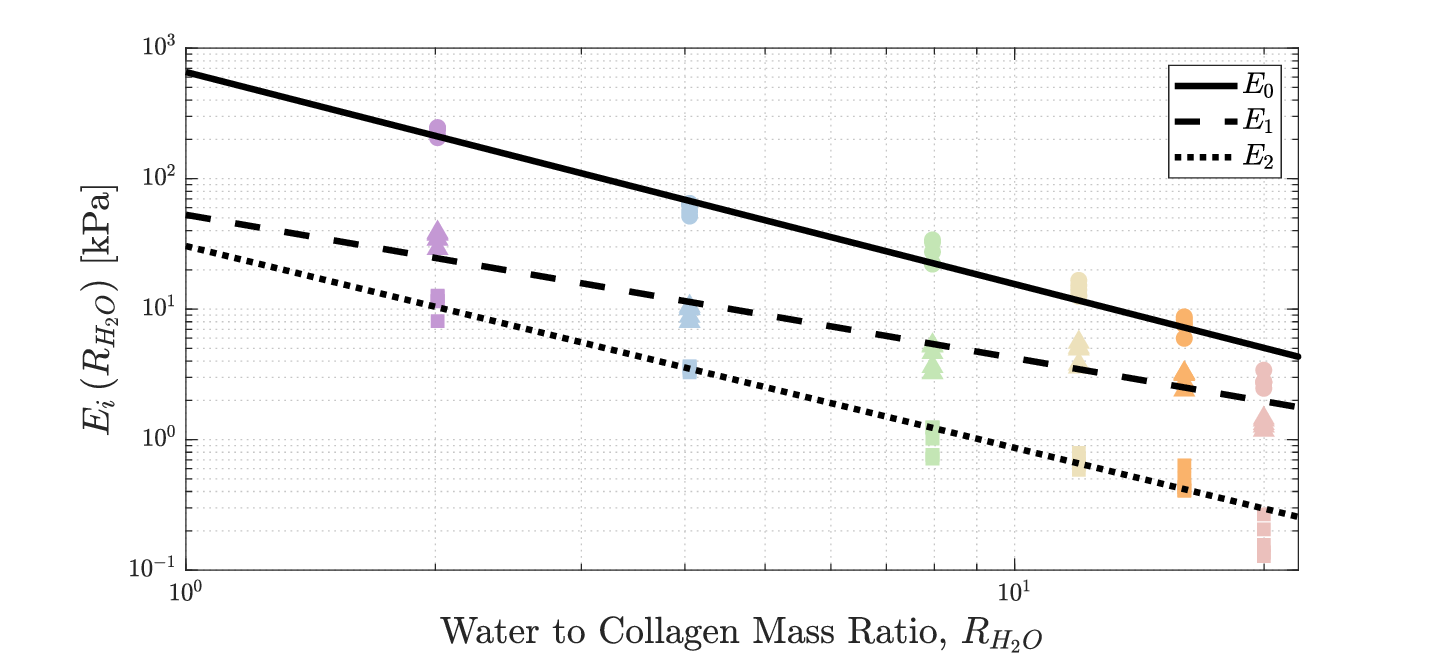}
\caption{Logarithmic plot of elastic moduli vs.\ water-to-gelatin powder mass ratio, with fits. The solid line represents $E_0$, the dashed line represents $E_1$, and the dotted line represents $E_2$. The moduli are well fit by a power law in the form (\ref{eq:moduli}). Colorful points represent experimental data as shown in Fig. \ref{fig:relaxation_modulus}}
\label{fig:moduli}
\end{figure}

Compared to literature, experimental stiffnesses were generally lower than the stiffnesses of collagen strands in Gautieri et al \cite{Cite5}. This is reasonable as a more organized strand structure would have a lower water ratio, and a much higher levels of crosslinking. Stiffnesses were also much higher than the pure gelatin hydrogels in Manish et al \cite{Cite8}, however, because of significant differences in manufacturing methods, there is reason to believe that a direct comparison would be invalid. Ballistic Gelatin is generally made in a ratio of 8 $< R_{H_2 O} <$ 10 in order to simulate human flesh \cite{Cite2}. NIH study found a young’s modulus for skeletal muscle of 24.6 ± 3.5 kPa \cite{Cite10}. Promisingly, using the power law relationship with parameters in table \ref{tab:moduli_fit_params} results in instantaneous stiffnesses ranging from 20 to 29 kPa. The fit resulting from our experimental study is consistent with water-to-collagen mass ratios commonly used in ballistics gelatin being used to simulate human flesh. 

\subsection{Viscoelastic Relaxation Times vs. Water-to-Collagen Mass Ratio}
\begin{figure}[htbp]
\centering
\includegraphics[width=\linewidth]{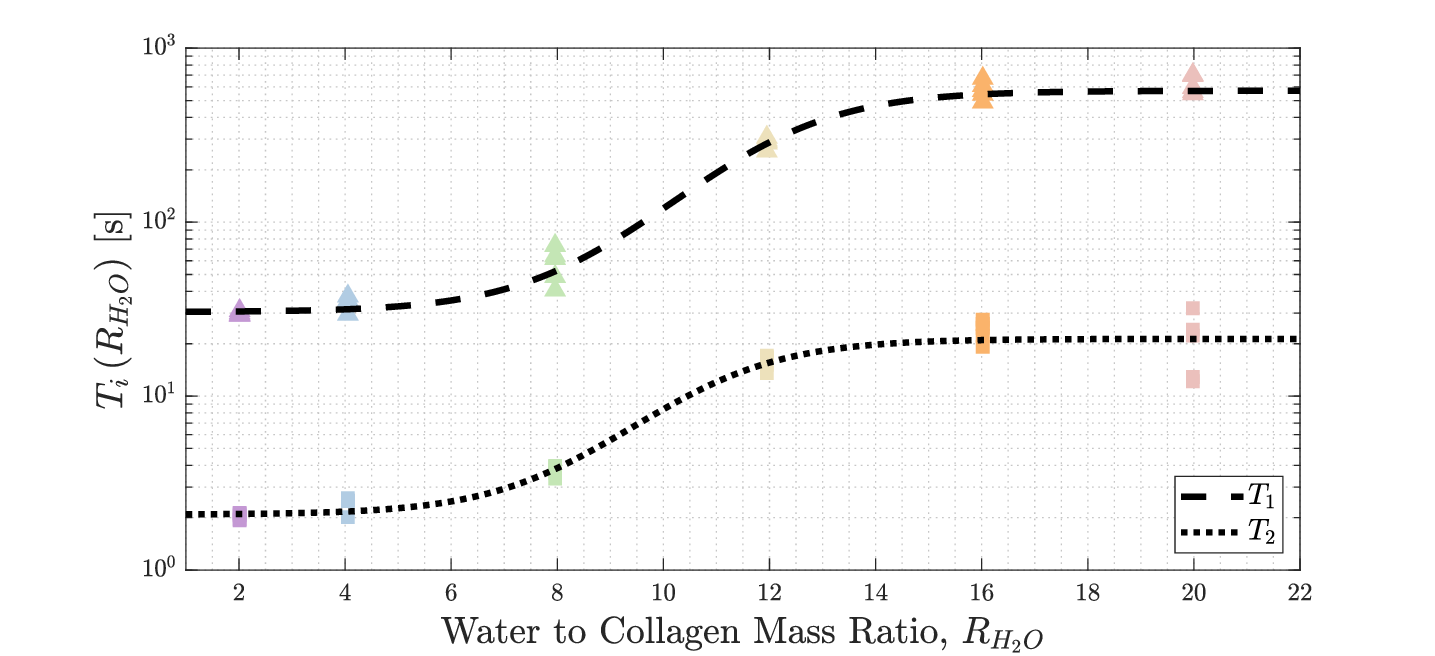}
\caption{Semi-log plot of the time constants vs.\ water-to-gelatin powder mass ratio, with fits. The dashed line represents $T_1$ and the dotted line represents $T_2$. The time constants are well fit via a sigmoid function in the form (\ref{eq:TimeConstants}) Colorful points represent experimental data as shown in Fig. \ref{fig:relaxation_modulus}.}
\label{fig:time_constants}
\end{figure}

Similarly, the time constants extracted from the step response fits can be plotted against the water mass ratio to reveal a clear relationship. As shown in Fig. \ref{fig:time_constants}, the time constant relationships are well fit via a sigmoid function of the form:
\begin{equation}\label{eq:TimeConstants}
    T_i(R_{H_2O})=\frac{c_i}{1+\exp[-d_i(R_{H_2O}-f_i)]}+g_i
\end{equation}
The same method\cite{Cite9} is again used to perform a weighted fit in order to account for large differences in absolute error among the different test conditions. Table \ref{tab:time_constant_fit_params} contains the values of the resulting parameters for both time constants. The time constant data was very consistent with a sigmoid function and therefore should be fairly robust in approximating the time constant of gelatin samples made. It can be hypothesized that at low water ratios the viscous time constant is dominated by the friction between and untangling of the collagen chains and at high water ratios, the time scale is dominated by the nanoporoelasticity of water moving through the polymer network.

\begin{table}[htbp]
\centering
\caption{Fit parameters of time constants vs.\ concentration. $c+g$ describes the maximum value of the time constant, $g$ describes the minimum value of the time constant, $d$ describes the steepness of the curve, and $f$ describes the mass-ratio at the midpoint of the sigmoid curve.}
\label{tab:time_constant_fit_params}
\begin{tabular}{|c||c|c|c|c|}
\hline
\textbf{Time Constant} & $\boldsymbol{c}$ \textbf{[s]} & $\boldsymbol{d}$ \textbf{[-]} & $\boldsymbol{f}$ \textbf{[-]} & $\boldsymbol{g}$ \textbf{[s]} \\
\hline
\hline
$\boldsymbol{T_1}$ & $539 \pm 43$ & $0.762 \pm 0.140$ & $12.1 \pm 0.3$ & $30.4 \pm 2.7$ \\
\hline
$\boldsymbol{T_2}$ & $19.2 \pm 2.1$ & $0.786 \pm 0.170$ & $10.9 \pm 0.6$ & $2.08 \pm 0.19$ \\
\hline
\end{tabular}
\end{table}

A 2011 study on collagen fibrils found relaxation times of 7 ± 2 s, and 102 ± 5 s for the short and long timescales respectively, which are quite similar to values expected from this fit for low water ratio gelatin \cite{Cite6}. A 2021 study using machine learning methods to fit indentation test data focused on only the longer of the two viscoelastic timescales and found it to be approximately 131 seconds \cite{Cite7}.

\subsection{Major Sources of Error}
\subsubsection{Quasi-Step-Response Underestimation}
To extract the viscoelastic properties of the samples, (\ref{eq:MaxWei}) was fit to the experimental data. However, (\ref{eq:MaxWei}) assumes a strain of amplitude $\varepsilon_0$ is applied instantaneously (at an infinitely fast rate). In actuality, the stress was applied at a finite rate of 1 mm/s. There is an analytical form for the quasi-static step response, but it was not used for the fit, as it has too many degrees of freedom to fit effectively. Both responses reach the same value as time increases, but there will be some error in the peak instantaneous relaxation modulus. This error can be estimated and increases as $T_1$ and $T_2$ decrease. Therefore, the maximum error due to the quasi-step-response should occur at the lowest value of water ratio, where the time constants are the lowest. Fig. \ref{fig:instantaneous_response} contains a plot of the response for the estimated values at the 2 to 1 ratio. The quasi-step-response undershoots the step response by 2.6\%. Because the error bounds on the instantaneous modulus are much greater than this error (±11\%), we will consider this error negligible. 

\subsubsection{Temperature Dependence of Parameters}
\begin{figure}[htbp]
\centering
\begin{subfigure}[t]{0.48\linewidth}
    \centering
    \includegraphics[width=\linewidth]{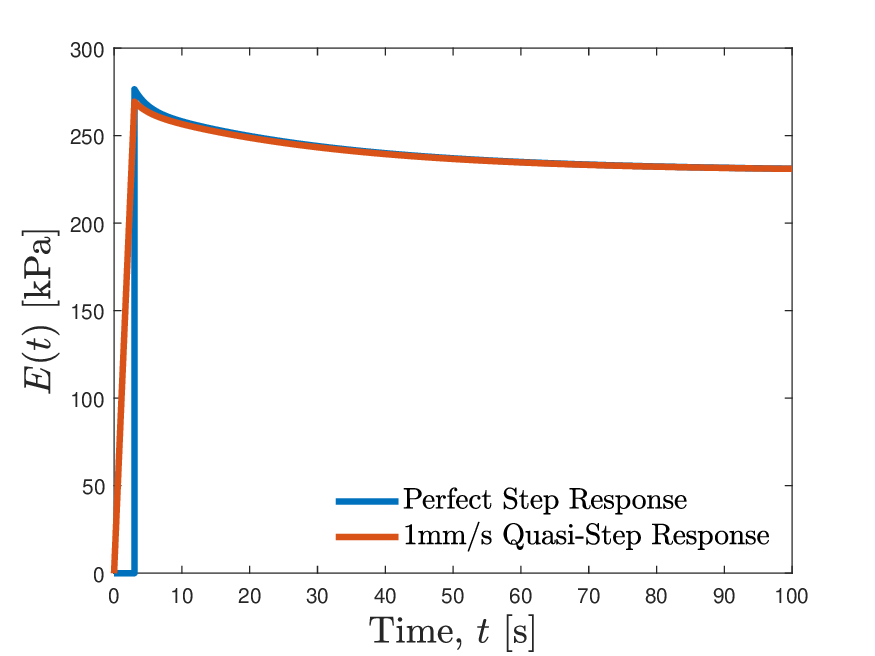}
    \caption{}
    \label{fig:instantaneous_response}
\end{subfigure}
\hfill
\begin{subfigure}[t]{0.48\linewidth}
    \centering
    \includegraphics[width=\linewidth]{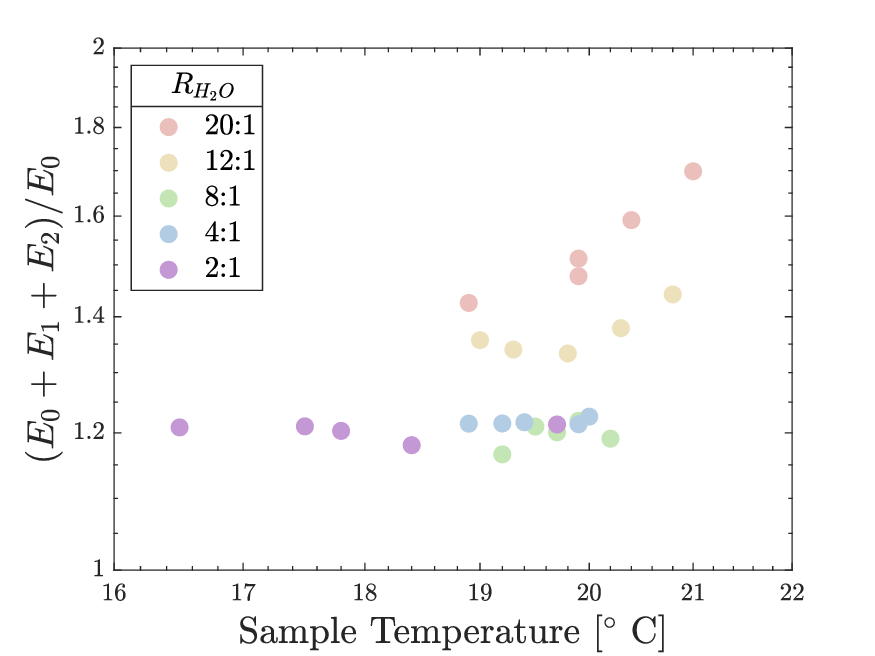}
    \caption{}
    \label{fig:temperature_dependence}
\end{subfigure}

\caption{(a) Plot of a perfect step response and a quasi-step response both based on the worst-case scenario parameters (shortest relaxation times) extracted from experimental data. The resulting overshoot is 2.6\%. 
(b) Plot of the normalized instantaneous moduli of each sample vs.\ the temperature of that sample at the time of testing.}
\label{fig:instantaneous_and_temperature}
\end{figure}

Gelatin stiffness can vary widely by temperature. Therefore, it is very important that the effect of temperature is isolated from the results of any experiment (or at the very least is determined to be inconsequential to the qualitative take aways of the experimental results). No pattern is immediately clear, however, if the two-time dependent moduli are normalized by the quasi-static relaxation modulus (as seen in Fig. \ref{fig:temperature_dependence}) a trend becomes clear. This variation, may also explain some of the deviations in $E_1$ from a simple power law. This variation is significant and warrants future investigation, however, Fig. \ref{fig:relaxation_modulus} shows plots of tests in a tight temperature range, and it becomes evident that the general trends in elastic moduli are consistent with the data if temperature is controlled. Additionally, the time constant data was extremely resistant to changes in temperature and showed no clear trends with temperature. 

\subsection{Future Investigations}
Further experimentation and fitting efforts may help to provide a fit that can predict the limits in the ratio between the instantaneous and quasi-static relaxation moduli. Additionally, these experiments provide room for future mathematical modeling to understand the physical phenomena behind these trends in both modulus and time constants. Little literature exists that points to trends like this, and this may be partially explained by the arbitrary choice to use water ratio instead of powder or water fraction, which seems to have worked quite well. 
Additionally, experiments performed by varying temperature in a controlled manner may allow for a better understanding of how temperature affects the viscoelastic properties of the material. In turn this will allow for the isolation of water ratio effects from temperature and allow for the use of the resulting fits to manufacture samples for use in various temperatures with specific stiffnesses. 

\section{Conclusions}
The compression experiments performed on the texture analyzer for samples of varying water to gelatin powder mass-ratios from 2:1 to 20:1 yielded results well fit to a Maxwell-Weichart linear-viscoelastic step-response of the form (\ref{eq:MaxWei}) with one purely elastic branch and two viscoelastic branches: one with a long and one with a short time scale. 

The relationship between the elastic moduli and the mass-ratio is well described via a decreasing power law (\ref{eq:moduli}). The relationship between the time constants and the mass-ratio is well described via an increasing sigmoid function (\ref{eq:TimeConstants}).

The relationship revealed between the viscoelastic material properties of gelatin and the water to gelatin powder mass-ratio of a sample through this simple set of experiments offer a strong basis for manufacturing gelatin samples with the desired mechanical properties to simulate a specific type of tissue with stiffnesses in this range. These experiments serve as a solid foundation for the use of gelatin as a cheap and safe representative hydrogel for use in soft-matter mechanics.

\begin{acks}
Thank you Dr. Steven Gillmer, Dr. Barbara Hughey, and Juergen Schoenstein for the support, advice, and guidance in running my experiments, analyzing my data, writing this paper, and of course, for being so generous with extensions! Thank you also to Professor Tal Cohen for helping me keep my models grounded in reality and discussing this project during our meetings!
\end{acks}

\bibliographystyle{SageV}
\bibliography{Main.bib}

@preamble{ "\newcommand{\noopsort}[1]{} "
	# "\newcommand{\printfirst}[2]{#1} "
	# "\newcommand{\singleletter}[1]{#1} "
	# "\newcommand{\switchargs}[2]{#2#1} " }

@article{Cite7, title={Identification of the concentration‐dependent viscoelastic constitutive parameters of gelatin by combining computational mechanics and machine learning}, volume={21}, DOI={https://doi.org/10.1002/pamm.202100250}, number={1}, journal={PAMM}, publisher={Wiley}, author={Abdolazizi, Kian P and Linka, Kevin and Sprenger, Johanna and Neidhardt, Maximilian and Schlaefer, Alexander and Cyron, Christian J}, year={2021}, month={Dec} }

@article{Cite2, title={Ballistic Gelatin Characterization and Constitutive Modeling}, ISBN={9781461402152}, DOI={https://doi.org/10.1007/978-1-4614-0216-9_7}, journal={Dynamic Behavior of Materials, Volume 1}, author={Cronin, D. S.}, year={2011}, pages={51–55} }

@article{Cite1, title={Characterization of 10% Ballistic Gelatin to Evaluate Temperature, Aging and Strain Rate Effects}, volume={51}, DOI={https://doi.org/10.1007/s11340-010-9438-z}, number={7}, journal={Experimental Mechanics}, author={Cronin, D. S. and Falzon, C.}, year={2010}, month={Nov}, pages={1197–1206} }

@article{Cite5, title={Modeling and measuring visco-elastic properties: From collagen molecules to collagen fibrils}, volume={56}, DOI={https://doi.org/10.1016/j.ijnonlinmec.2013.03.012}, journal={International Journal of Non-Linear Mechanics}, author={Gautieri, Alfonso and Vesentini, Simone and Redaelli, Alberto and Ballarini, Roberto}, year={2013}, month={Nov}, pages={25–33} }

@article{Cite4, title={Viscoelasticity and poroelasticity in elastomeric gels}, volume={25}, DOI={https://doi.org/10.1016/s0894-9166(12)60039-1}, number={5}, journal={Acta Mechanica Solida Sinica}, author={Hu, Yuhang and Suo, Zhigang}, year={2012}, month={Oct}, pages={441–458} }

@article{Cite8, title={Influence of water content on the mechanical behavior of gelatin based hydrogels: Synthesis, characterization, and modeling}, volume={233}, DOI={https://doi.org/10.1016/j.ijsolstr.2021.111219}, journal={International Journal of Solids and Structures}, author={Manish, Vivek and Arockiarajan, A. and Tamadapu, Ganesh}, year={2021}, month={Dec}, pages={111219} }

@article{Cite10, title={Endothelial, cardiac muscle and skeletal muscle exhibit different viscous and elastic properties as determined by atomic force microscopy}, volume={34}, DOI={https://doi.org/10.1016/S0021-9290(01)00149-X}, number={12}, journal={Journal of Biomechanics}, author={Mathur, A B and Collinsworth, A M and Reichert, W M and Kraus, W E and Truskey, G A}, year={2001}, month={Dec}, pages={1545–1553} }

@misc{Cite3, title={ENGINEERING VISCOELASTICITY}, url={https://web.mit.edu/course/3/3.11/www/modules/visco.pdf}, institution={MIT}, author={Roylance, David}, year={2001} }

@article{Cite6, title={Viscoelastic Properties of Isolated Collagen Fibrils}, volume={100}, DOI={https://doi.org/10.1016/j.bpj.2011.04.052}, number={12}, journal={Biophysical Journal}, author={Shen, Zhilei Liu and Kahn, Harold and Ballarini, Roberto and Eppell, Steven J.}, year={2011}, month={Jun}, pages={3008–3015} }

@misc{Cite9,  title = {NIST Engineering Statistics Handbook}, url={https://www.itl.nist.gov/div898/handbook/pmd/section4/pmd452.htm.}, publisher={NIST}, year={2015}, month={Apr} , chapter={4.4.5.2}}

\end{document}